\documentclass[useAMS,usenatbib]{mn2e}
\usepackage[dvipdfmx]{graphicx}
\usepackage[dvipdfmx]{color}
\usepackage{multirow}
\def\aj{AJ}             	
\def\apj{ApJ}           	
\def\apjl{ApJ}          	
\def\aap{A\&A}          	
\def\mnras{MNRAS}       	
\def\na{New Astron.}        	
\def\jcap{JCAP}   

\def\gsim{\hspace{0.3em}\raisebox{0.4ex}{$>$}\hspace{-0.75em}\raisebox{-.7ex}{$\sim$}\hspace{0.3em}}
\def\lsim{\hspace{0.3em}\raisebox{0.4ex}{$<$}\hspace{-0.75em}\raisebox{-.7ex}{$\sim$}\hspace{0.3em}}

\title[Emergence of a stellar cusp by dynamical friction]{Emergence of a stellar cusp by a dark matter cusp in a low-mass compact ultra-faint dwarf galaxy}
\author[S. Inoue]
{\parbox[t]{\textwidth} 
{Shigeki Inoue\thanks{E-mail: shigeki.inoue@ipmu.jp}}
\\ \\
Kavli Institute for the Physics and Mathematics of the Universe (WPI), UTIAS, The University of Tokyo, Chiba 277-8583, Japan\\
Department of Physics, School of Science, The University of Tokyo, Bunkyo, Tokyo 113-0033, Japan
}

\begin{document}

\pagerange{\pageref{firstpage}--\pageref{lastpage}} \pubyear{2014}

\maketitle

\label{firstpage}

\begin{abstract}
  Recent observations have been discovering new ultra-faint dwarf galaxies as small as $\sim20~{\rm pc}$ in half-light radius and $\sim3~{\rm km~s^{-1}}$ in line-of-sight velocity dispersion. In these galaxies, dynamical friction on a star against dark matter can be significant and alter their stellar density distribution. The effect can strongly depend on a central density profile of dark matter, i.e. cusp or core. In this study, I perform computations using a classical and a modern analytic formulae and $N$-body simulations to study how dynamical friction changes a stellar density profile and how different it is between a cuspy and a cored dark matter haloes. This study shows that, if a dark matter halo has a cusp, dynamical friction can cause shrivelling instability which results in emergence of a stellar cusp in the central region $\lsim2~{\rm pc}$. On the other hand, if it has a constant-density core, dynamical friction is significantly weaker and does not generate a stellar cusp even if the galaxy has the same line-of-sight velocity dispersion. In such a compact and low-mass galaxy, since the shrivelling instability by dynamical friction is inevitable if it has a dark matter cusp, absence of a stellar cusp implies that the galaxy has a dark-matter core. I expect that this could be used to diagnose a dark matter density profile in these compact ultra-faint dwarf galaxies.

\end{abstract}


\section{Introduction}
\label{Intro}
Dark matter (DM) density profiles in dwarf galaxies have long been debated. Theoretical studies such as cosmological $N$-body simulations have demonstrated that DM density increases toward the galactic centre independent of a halo mass \citep[e.g.][]{dc:91,nfw:97,kkb:01,swv:08,imp:11}. On the other hand, observations proposed that dwarf galaxies seem to have nearly constant densities of DM at their central regions \citep[`cusp/core problem', e.g.][]{gww:07,obb:10,hc:12}.\footnote{Low surface brightness galaxies have also been observed to have DM cores \citep[][and references therein]{d:10} although I do not discuss these galaxies.}

As a possible solution, if DM haloes consist of warm or self-interacting particles, all dwarf galaxies are expected to have central DM cores. It has also been proposed, alternatively, that (recursive) baryonic feedback can turn a cusp into a core by flattening the inner slopes of the primordial DM density profiles in dwarf galaxies as massive as $M_{\rm DM}\sim 10^{10}$--$10^{11}~{\rm M_\odot}$ \citep[e.g.][]{g:10,pg:12,om:14,ewg:16,dbd:16}. If the latter scenario is the case, since dynamical masses of some ultra-faint dwarf galaxies (UFDs) in the local group have been observed to be significantly smaller than the mass threshold above which the baryonic effect is influential to their central DM densities, they could be expected to preserve the primordial DM density profiles which may be cuspy. Accordingly, it is interesting to try to determine DM density profiles of such low-mass UFDs. It is, however, still impossible to know whether their DM haloes have cusps or cores because of only a handful of stars observable by spectroscopy to measure their line-of-sight velocities (LOSVs) and model their DM haloes. Hence, it is worthwhile looking for an alternative method to deduce which type of DM the low-mass galaxies have, cusp or core. For example, \citet{plc:16} have proposed a method using a fraction of wide binaries which can be disrupted by tidal force depending on their DM potential in UFDs.

The smallest UFDs are as tiny as $R_{\rm h}\lsim30~{\rm pc}$ in half-light radius and $L \sim10^{2-3}~{\rm L_\odot}$ in luminosity \citep[e.g.][]{w:05,bel:09,l:15,m:16,hc:16,s:16} although current observations still cannot reject the possibility that some of them are extended globular clusters. Recently, \citet{h:16} has analytically discussed that dynamical friction (DF) on a star against dark matter may be marginally effective on the timescale of $\sim10~{\rm Gyr}$ in Draco II --- observed physical properties of which are $R_{\rm h}=19^{+8}_{-6}~{\rm pc}$, brightness $M_{\rm v}=-2.9\pm0.8$, LOSV dispersion $\sigma_{\rm h}=2.9\pm2.1~{\rm km~s^{-1}}$ measured within $R_{\rm h}$ \citep{m:16} --- by adopting the Chandrasekhar DF formula \citep{c:43} to his singular isothermal DM halo model. His result implies that DF against DM could significantly change stellar distribution in UFDs more compact and/or less massive than Draco II, which will be discovered by future observations.

The effect of DF strongly depends on the DM density profile. It has been known that DF drag force becomes significantly weaker in cored density distribution than in cuspy one, once a massive particle enters the core \citep[e.g.][]{hg:98}. Studies using $N$-body simulations have demonstrated that drag force by DF does cease practically in a constant-density core, probably by non-linear effects \citep[e.g.][]{gmr:06,rgm:06,i:09,i:11,ac:14,pgr:15,prg:16}. Therefore, if an extremely low-mass UFD has a constant-density core of DM, DF could be too weak to affect the stellar distribution. On the other hand, if such a UFD has a DM cusp, DF against DM could be strong enough to make alterations to its stellar distribution, such as emergence of a stellar cusp or formation of a nucleus cluster as a remnant of stars fallen into the galactic centre. Current observations of low-mass compact UFDs are limited to a close distance of $d\lsim30~{\rm kpc}$ from the sun because of their faintness. At this distance, each star in a UFD can be resolved since the typical size of observational smearing is smaller than the mean separation of stars even at the galactic centres. Therefore, the expected stellar cusp and  the nucleus cluster would be observed as a dense group of stars at the galactic centre if it exists.

This study addresses the effect of DF by DM on stellar distribution in an extremely low-mass and compact UFD and focus on how different it is between cuspy and cored DM density profiles. In Section \ref{analysis}, I perform analytical estimation of stellar shrivelling due to DF based on a classical and a modern DF formulae. In Section \ref{SimSec}, I perform $N$-body simulations resolving every single star and demonstrate the same as the analytical estimation presented in Section \ref{analysis}. Finally, I present discussion and summary of this work in Section \ref{Discussion}.

\section{Analysis using dynamical friction formulae}
\label{analysis}
In this study, I describe a density profile of a DM halo by a Dehnen model \citep{d:93},
\begin{equation}
  \rho_{\rm DM}(r) = \frac{\rho_{\rm DM,0}r_{\rm s}^4}{r^\gamma(r+r_{\rm s})^{4-\gamma}},
\label{density}
\end{equation}
where $\rho_{\rm DM,0}$ and $r_{\rm s}$ are scale density and radius, and $\gamma$ is an inner density slope. I assume a cuspy DM halo to be represented by setting $\gamma=1$, which corresponds to the Hernquist model profile \citep{h:90}, and a cored DM halo is represented by $\gamma=0$. Local velocity dispersion of DM is generally computed by solving Jeans equation, 
\begin{equation}
  \sigma_{\rm DM}^2(r) = \frac{1}{\rho_{\rm DM}}\int^\infty_r\rho_{\rm DM}\frac{GM_{\rm DM}(r')}{r'^2}~\textrm{d}r',
\label{Jeans}
\end{equation}
where $G$ is the gravitational constant, and $M_{\rm DM}(r)$ is mass of DM enclosed within $r$. Here, I assume that gravity of a baryon component is negligible and that velocity distribution is isotropic. The analytic solutions of $\sigma_{\rm DM}$ for $\gamma=0$ and $1$ can be found in \citet{d:93} and \citet{h:90}.

\begin{figure}
  \includegraphics[width=\hsize]{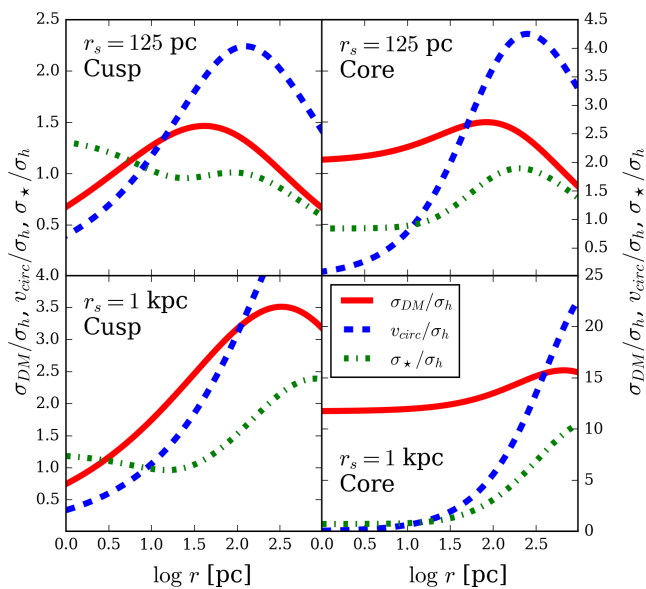}
  \caption{Radial profiles of local velocity dispersions of DM and stars, and circular velocities in the cuspy (left) and the cored (right) halo models with $r_{\rm s}=125~{\rm pc}$ (top) and $1~{\rm kpc}$ (bottom). The left and right ordinates are for the left and right panels. The profiles are normalised by $\sigma_{\rm h}$.}
  \label{velocities}
\end{figure}

This study discusses how the DF against DM affects stellar distribution and how different it is between cuspy and cored DM haloes. I use a Plummer's model for stellar distribution of a compact UFD,
\begin{equation}
  \rho_\star(r) = \frac{3M_\star}{4\pi r_\star^3}\left(1+\frac{r^2}{r_\star^2}\right)^{-\frac{5}{2}},
\label{star}
\end{equation}
where $M_\star$ and $r_\star$ are the total mass and scale radius of stars. This model has a core of stars in $r\ll r_\star$ where the density is nearly constant. Two-dimensional half-light radius $R_{\rm h}$ and integrated mass $M_{\rm h}$ inside $R_{\rm h}$ are obtained by assuming a constant mass-to-luminosity ratio and integrating equation (\ref{star}).  In this study, I assume $r_\star=20~{\rm pc}$ ($R_{\rm h}=r_\star$ in a Plummer's model). Luminosity-weighted LOSV dispersion inside $R_{\rm h}$ is given as 
\begin{equation}
  \sigma^2_{\rm h} = \frac{4\pi}{M_{\rm h}}\int^{\infty}_0dz\int^{R_{\rm h}}_0\rho_\star\sigma^2_{\star}(R',z)R'~\textrm{d}R',
\end{equation}
where stellar velocity dispersion $\sigma_{\star}$ is computed from equation (\ref{Jeans}) in which $\rho_\star$ is substituted for $\rho_{\rm DM}$. Since stellar gravity is now assumed to be negligible, setting $\sigma_{\rm h}$ gives $\rho_{\rm DM,0}$ when the other parameters in equation (\ref{density}) are fixed:\footnote{When $\sigma_{\rm h}=1.5~{\rm km~s^{-1}}$, the values of $\rho_{\rm DM,0}$ in the cored DM models are $9.8$, $7.0$, $5.8$ and $5.1\times10^{-1}~{\rm M_\odot~pc^{-3}}$ for $r_{\rm s}=125$, $250$, $500~{\rm pc}$ and $1~{\rm kpc}$. Those in the cuspy models are $1.1$, $0.45$, $0.20$ and $0.096\times10^{-1}~{\rm M_\odot~pc^{-3}}$, respectively.} $\sigma_{\rm h}^2\propto\rho_{\rm DM,0}$. In what follows, I discuss the two cases of $\sigma_{\rm h}$ to $1.5$ and $3.0~{\rm km~s^{-1}}$.\footnote{The cusp and the core models of equation (\ref{density}) have the finite total masses, $M_{\rm cusp,tot}=2\pi\rho_{\rm DM,0}r_{\rm s}^3$ and $M_{\rm core,tot}=(4/3)\pi\rho_{\rm DM,0}r_{\rm s}^3$, respectively. When $\sigma_{\rm h}=1.5~{\rm km~s^{-1}}$, for $r_{\rm s}=125~{\rm pc}$, the total masses of cuspy and cored haloes are $M_{\rm cusp,tot}=1.3\times10^6~{\rm M_\odot}$ and $M_{\rm core,tot}=8.0\times10^6~{\rm M_\odot}$. For $r_{\rm s}=1~{\rm kpc}$, $M_{\rm cusp,tot}=6.0\times10^7~{\rm M_\odot}$ and $M_{\rm core,tot}=2.2\times10^9~{\rm M_\odot}$.} Fig. \ref{velocities} illustrates radial profiles of $\sigma_{\rm DM}$, circular velocities $v_{\rm circ}\equiv\sqrt{GM_{\rm DM}(r)/r}$ and $\sigma_\star$ normalised by $\sigma_{\rm h}$ in my cuspy and cored halo models with $r_{\rm s}=125~{\rm pc}$ and $1~{\rm kpc}$.

\subsection{Analytic formulae of dynamical friction}
\subsubsection{The Chandrasekhar formula}
\begin{figure*}
  \includegraphics[width=\hsize]{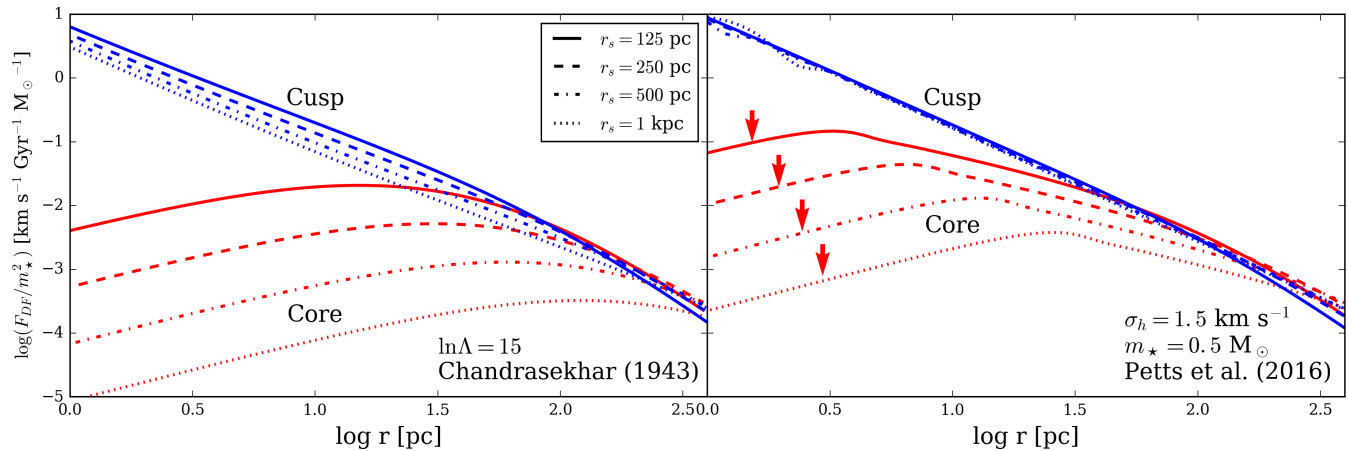}
  \caption{\textit{Left panel}: drag force by DF computed with equation (\ref{formula}) with $\ln\Lambda=15$ normalised by $m_\star^2$ in the halo model of equation (\ref{density}). The blue and red lines indicate the results in the cuspy and the cored DM models, respectively. \textit{Right panel}: same as the left panel but computed with equation (\ref{Petts_formula}) in the case of $\sigma_{\rm h}=1.5~{\rm km~s^{-1}}$ and $m_\star=0.5~{\rm M_\odot}$. The red arrows indicate tidal-stalling radii, $r_{\rm TS}$, in the cored DM haloes. In the cuspy haloes, $r_{\rm TS}<0.1~{\rm pc}$.}
  \label{CF_comp}
\end{figure*}
\label{classical}
I consider DF against DM on a star. The Chandrasekhar DF formula under Maxwellian velocity distribution\footnote{Although \citet{c:43} has also proposed more general forms of analytic DF not relying on Maxwell distribution, I refer to equation (\ref{formula}) as Chandrasekhar formula in this paper.} is given as 
\begin{equation}
  F_{\rm DF} = -\frac{4\pi\ln\Lambda G^2\rho_{\rm DM}m_\star^2}{v_\star^2}\left[{\rm erf}(X)-\frac{2X}{\sqrt{\pi}}\exp(-X^2)\right],
\label{formula}
\end{equation}
where $m_\star$ and $v_\star$ are a mass and a velocity of a star, $X\equiv v_\star/(\sqrt{2}\sigma_{\rm DM})$, and $\Lambda$ is a parameter, whose proper value is still under debate \citep[e.g.][]{ac:14,jp:05,pgr:15,prg:16}\footnote{Basically, $\Lambda$ is defined to be a ratio between the minimum and maximum impact parameters of two-body gravitational interaction, i.e., $\Lambda\equiv b_{\rm max}/b_{\rm min}$, where $b_{\rm min}\sim Gm_\star/v_\star^2$, $b_{\rm max}\sim r_{\rm s}$ in the classical formula \citep{c:43,bt:08}.}. The direction of $\mathbfit F_{\rm DF}$ is presumed to be opposite to the velocity vector $\mathbfit v_\star$. By assuming a circular orbit, i.e. $v_\star=v_{\rm circ}$, equation (\ref{formula}) can be solved with $\sigma_{\rm DM}$ from equation (\ref{Jeans}). In the case considered here, equation (\ref{formula}) is independent of $\rho_{\rm DM,0}$ (i.e. $\sigma_{\rm h}$) since $\rho_{\rm DM}$, $v_\star^2$ and $\sigma_{\rm DM}^2$ are proportional to $\rho_{\rm DM,0}$ although the DF timescale, $\sim m_\star\sigma_{\rm h}/F_{\rm DF}$, depends on $\rho_{\rm DM,0}$. In addition, because the parentheses in equation (\ref{formula}) is independent of $m_\star$, $F_{\rm DF}/m_\star^2$ is independent of $m_\star$.

The left panel of Fig. \ref{CF_comp} shows the Chandrasekhar DF force of equation (\ref{formula}) with $\ln\Lambda=15$ normalised by $m_\star^2$ in the cuspy and the cored haloes with various $r_{\rm s}$. The Figure indicates that the strength of DF is remarkably different between the cuspy and the cored DM haloes; DF increases monotonically towards the centre in a DM cusp (the blue lines), whereas it is approximately constant or gently decreases in a core (the red lines). The behaviour of DF is almost independent of $r_{\rm s}$ in the cuspy haloes, whereas DF in a core becomes weaker when $r_{\rm s}$ is larger. Although the difference of DF inside $r_{\rm s}$ between the cusp and the core becomes smaller with decreasing $r_{\rm s}$, it is still quite large in $r\lsim10~{\rm pc}$ even in the case of $r_{\rm s}=125~{\rm pc}$. This means that, if a stellar component of a UFD is deeply embedded in a DM halo (i.e. $R_{\rm h}\ll r_{\rm s}$), one can expect that DF strongly depends on a density profile of DM and could change the stellar distribution if the compact UFD has a cusp of DM. Moreover, in a cuspy halo, a star undergoing DF migrates into an inner radius, at which DF is even stronger (see Section \ref{interp}).

\subsubsection{Petts et al. formula}
\label{modern}
The Chandrasekhar formula of equation (\ref{formula}) is, however, based on several assumptions such as the Maxwellian velocity distribution and the invariable parameter of $\Lambda$. Therefore, improved formulae have been invented by previous studies. Recently, \citet{prg:16} proposed more sophisticated DF modelling based of the general Chandrasekhar formula \citep[equations 25 and 26 in][]{c:43}, which uses a distribution function instead of the Maxwellian distribution and takes high-velocity encounters into account. They demonstrated that their improved DF model can reproduce orbits of infalling particles in cuspy and cored density fields better than the classical formula. They formulate DF force\footnote{\citet{prg:16} proposed two models of DF: `P16' and `P16f'. I use their P16f model in this study since they concluded that P16f is more accurate than P16.} as 
\begin{equation}
  F_{\rm DF} = -\frac{2\pi^2 G^2\rho_{\rm DM}m_\star^2}{v_\star^2}\int^{v_{\rm esc}}_0J(v_{\rm DM})f(v_{\rm DM})v_{\rm DM}~\textrm{d}v_{\rm DM},
\label{Petts_formula}
\end{equation}

\begin{equation}
  J= \int^{v_\star+v_{\rm DM}}_{|v_\star-v_{\rm DM}|}\left(1+\frac{v_\star^2-v_{\rm DM}^2}{V^2}\right)\log\left(1+\frac{b_{\rm max}^2V^4}{G^2m_\star^2}\right)~\textrm{d}V,
\label{Petts_J}
\end{equation}
where $f(v_{\rm DM})$ represents a distribution function of DM, which is defined so that $4\pi \int f(v_{\rm DM})v_{\rm DM}^2~\textrm{d}v_{\rm DM}=1$, and escape velocity $v_{\rm esc}=\sqrt{-2\Phi}$, and $V$ corresponds to relative velocity of encounter. In equation (\ref{Petts_J}), the maximum impact parameter
\begin{equation}
  b_{\rm max} = \min \left( \frac{\rho_{\rm DM}(r)}{\textrm{d}\rho_{\rm DM}/\textrm{d}r},~r\right).
\label{bmax}
\end{equation}
The value of $\rho_{\rm DM}/(\textrm{d}\rho_{\rm DM}/\textrm{d}r)$ can be taken as the distance within which the density field can be considered to be homogeneous \citep{jp:05,jkb:11}. However, since it can diverge in a constant-density core, $b_{\rm max}$ is limited to be $\leq r$ \citep{pgr:15}. Equation (\ref{Petts_J}) can be solved analytically (see Appendix \ref{anaJ}). 

In addition, \citet{prg:16} also introduced a `tidal-stalling' radius, $r_{\rm TS}$, at which the tidal radius of a massive particle is equal to its orbital radius. The tidal radius is 
\begin{equation}
  r_{\rm t} = \frac{Gm_\star}{\Omega^2-\textrm{d}^2\Phi/\textrm{d}r^2},
\label{tidalstallingradius}
\end{equation}
where $\Omega^2=GM_{\rm DM}/r^3$. They argued that DF ceases within the tidal-stalling radius because of non-linear effects, and showed that the radius of $r_{\rm TS}$ matches well results of $N$-body simulations (a radius of DF cessation in a core, see Section \ref{Intro}). In the \citeauthor{prg:16} DF model, $F_{\rm DF}=0$ in $r<r_{\rm TS}$ although equation (\ref{Petts_formula}) still returns a non-zero value.  

The right panel of Fig. \ref{CF_comp} shows the \citeauthor{prg:16} DF force of equation (\ref{Petts_formula}) normalised by $m_\star^2$. Unlike the Chandrasekhar formula, now $F_{\rm DF}/m_\star^2$ weakly depends on $\sigma_{\rm h}$ and $m_\star$. Here, I assume $\sigma_{\rm h}=1.5~{\rm km~s^{-1}}$ and $m_\star=0.5~{\rm M_\odot}$, however the results hardly change between $\sigma_{\rm h}=1.5$ and $3.0~{\rm km~s^{-1}}$. Radii of $r_{\rm TS}$ become about $1.4$ times smaller when $\sigma_{\rm h}=3.0~{\rm km~s^{-1}}$. In the right panel of Fig. \ref{CF_comp}, although the differences between the cuspy and the cored haloes are still quite large, it is remarkable that the \citeauthor{prg:16} formula predicts DF significantly stronger than the classical formula in the central regions of the cored haloes \citep[`super-Chandrasekhar DF',][]{rgm:06,gmr:06,zk:16,prg:16}. On the other hand, the DF in the cuspy haloes is similar to that given by the Chandrasekhar formula.

\subsection{Orbital integration with the formulae}
\label{orbitinteg}
Using the models and the DF formulae described above, I perform orbital integration of stars under the potential given by the DM distribution of equation (\ref{density}). With the initial spatial distribution of equation (\ref{star}), the velocity distribution of stars is given by Eddington's formula \citep{bt:08} with isotropy. I do not take into account mutual interactions between the stars, therefore the result is independent of the number of stars. For the sake of statistics, I use a random sample of ten million stars in each run. While integrating their orbits with respect to time, the stars are decelerated by DF represented by the analytic formulae every timestep. I assume $m_\star=0.5~{\rm M_\odot}$ as a typical mass of a star as old as $\sim10~{\rm Gyr}$ \citep{k:02,m:05}. The analytic DF is considered to work until a star reaches the radius $r_{\rm limit}$ at which $M_{\rm DM}(r_{\rm limit})=m_\star$. In my models, $r_{\rm limit}\simeq0.1$ and $0.5~{\rm pc}$ in the cuspy and the cored DM haloes. When a star enters $r_{\rm limit}$ with a velocity slower than $v_{\rm circ}|_{r=r_{\rm limit}}$, the star is stopped there and considered to be fallen into the galactic centre by DF. When the Chandrasekhar formula is applied, I set $\ln\Lambda=15$. When the \citeauthor{prg:16} formula is applied, DF ceases within $r_{\rm TS}$ (i.e. $F_{\rm DF}=0$). I use a second-order leap-frog integrator for the orbital computations with a constant and shared timestep of $\Delta t=0.01\times r_{\rm limit}/v_{\rm circ}|_{r=r_{\rm limit}}$. I confirmed the convergence of my results with respect to $\Delta t$. Orbits of stars are time-integrated until $t=10~{\rm Gyr}$, and I obtain stellar surface densities as functions of radius in the runs.

\begin{figure}
  \includegraphics[width=\hsize]{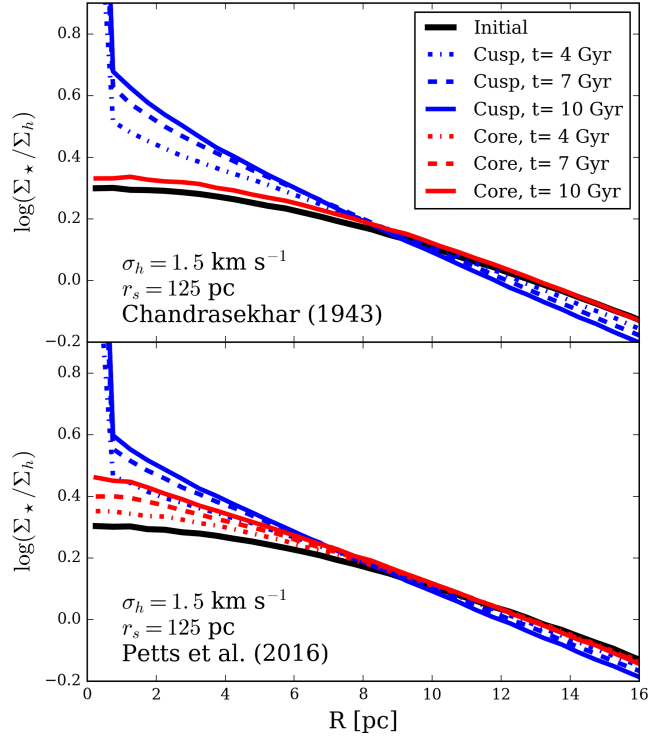}
  \caption{Time-evolution of the stellar surface density profiles in the cuspy and the cored DM density model with $r_{\rm s}=125~{\rm pc}$ and $\sigma_{\rm h}=1.5~{\rm km~s^{-1}}$. In the top and bottom panels, equations (\ref{formula}) and (\ref{Petts_formula}) are used to model DF effect. In the top panel, results at $t=4$ and $7~{\rm Gyr}$ in the cored halo are not shown. The black solid line indicates the initial state of the profile. The ordinates are normalised by $\Sigma_{\rm h}\equiv M_\star/(2\pi R_{\rm h}^2)$. The density peaks at the centre reach $\log(\Sigma_\star/\Sigma_{\rm h})=2.2$ and $2.1$ at $t=10~{\rm Gyr}$ in the top and bottom panels.}
  \label{OI_Hernq}
\end{figure}
Fig. \ref{OI_Hernq} shows the results of the orbital integration using the Chandrasekhar (top) and \citeauthor{prg:16} (bottom) formulae in the cuspy and the cored haloes with $r_{\rm s}=125~{\rm pc}$ and $\sigma_{\rm h}=1.5~{\rm km~s^{-1}}$. Although the stellar surface density is nearly constant in $R\lsim5~{\rm pc}$ in the initial state (black line), the density profiles are significantly steepened after $t\sim4~{\rm Gyr}$ in the cuspy DM halo (blue lines). In addition, sharp stellar cusps like nucleus clusters emerge at the galactic centres, which have five and four per cents of the total number of stellar particles within $R<0.5~{\rm pc}$ in the top and the bottom panels. The cusps mainly consist of stars fallen into the centres by DF. The steepened stellar distribution profiles are nearly exponential outside the stellar cusps. Because the Chandrasekhar and the \citeauthor{prg:16} formulae are not significantly different (Fig. \ref{CF_comp}), the results of Fig. \ref{OI_Hernq} are similar in the cuspy halo. These results corroborate the expectation that, as \citet{h:16} proposed,  a stellar distribution in a low-mass compact UFD can be affected by DF against DM if it has a cusp. 

If a DM halo has a core, however, the DF approximated by the analytic formulae is significantly less efficient to steepen the stellar profile  (the red lines), in spite of the same $\sigma_{\rm h}$ meaning similar DM masses within $R_{\rm h}$. When the Chandrasekhar formula is applied (the top panel), the stellar density hardly changes even at $t=10~{\rm Gyr}$ in the cored DM halo. Although the \citeauthor{prg:16} formula (the bottom panel) steepens the stellar density profile more than the classical formula, the density slope is clearly shallower than that in the cuspy DM, and a stellar cusp does not form. No stars are fallen into the centre by either DF modellings. The absence of the stellar cusp is due to the week DF in the DM core and the DF cessation assumed in $r<r_{\rm TS}$ in the \citeauthor{prg:16} model.

\begin{figure*}
  \begin{minipage}{\hsize}
    \includegraphics[width=\hsize]{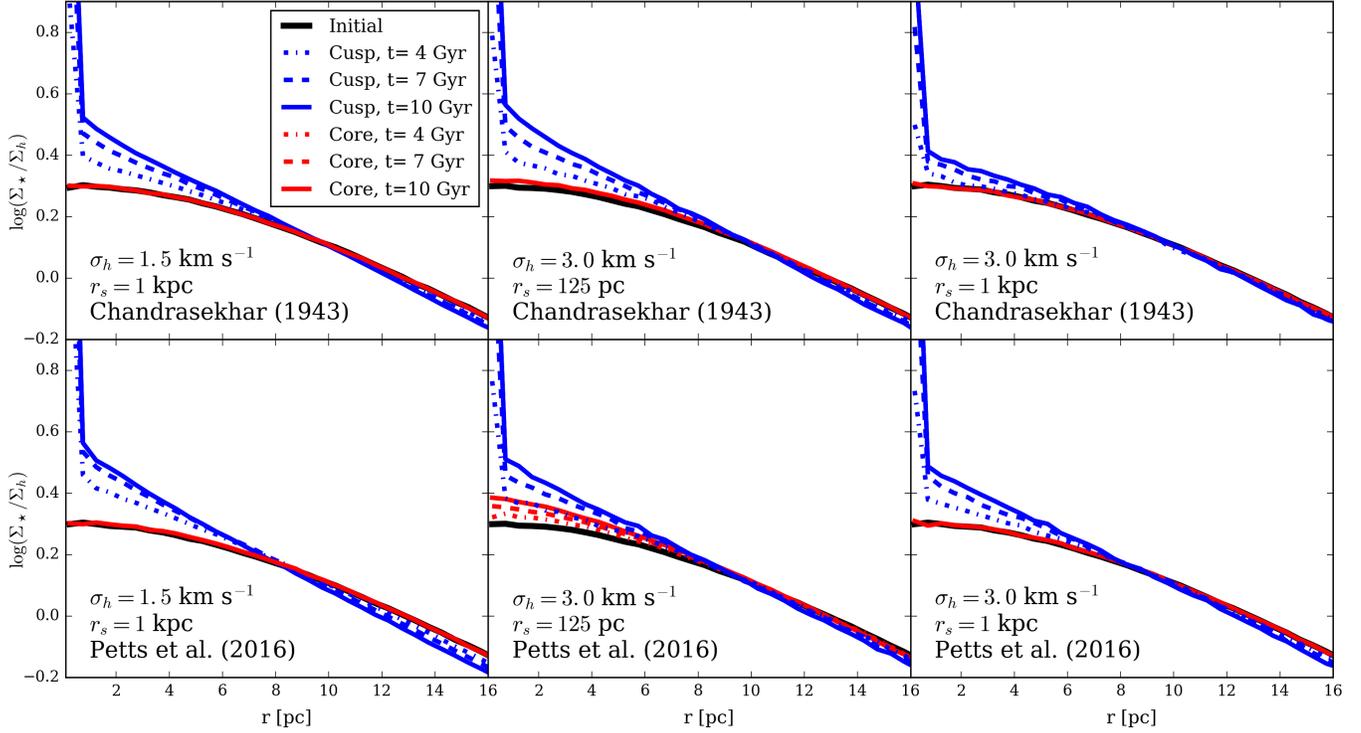}
    \caption{Same as Fig. \ref{OI_Hernq} but for halo models different in $r_{\rm s}$ and $\sigma_{\rm h}$. In the top panels, the density peaks at the centre reach $\log(\Sigma_\star/\Sigma_{\rm h})=1.6$, $1.6$ and $1.1$ at $t=10~{\rm Gyr}$ from left to right. In the bottom panels, the density peaks are $\log(\Sigma_\star/\Sigma_{\rm h})=2.1$, $1.6$ and $1.6$, respectively.}
    \label{OI_others}
  \end{minipage}
\end{figure*}
Fig. \ref{OI_others} shows the same results but for different settings for  $r_{\rm s}$ and $\sigma_{\rm h}$. The effect of DF becomes weaker for larger $r_{\rm s}$ and $\sigma_{\rm h}$ (i.e. higher $\rho_{\rm DM,0}$), but the steepening of a surface density profile and formation of a stellar cusp by a DM cusp can be seen even when $r_{\rm s}=1~{\rm kpc}$ and $\sigma_{\rm h}=3.0~{\rm km~s^{-1}}$. In the case of cored DM haloes, on the other hand, the stellar density profiles are almost intact even though $\sigma_{\rm h}$ and $r_{\rm s}$ are the same as in the cuspy halo models. In the case of the cored halo with $\sigma_{\rm h}=3.0~{\rm km~s^{-1}}$ and $r_{\rm s}=125~{\rm pc}$, the \citeauthor{prg:16} formula predicts weak steepening, but a stellar cusp does not emerge. Thus, significance of the DF effect strongly depends on whether the DM halo has a cusp or a core even if the $r_{\rm s}$ and $\sigma_{\rm h}$ are the same. The most noticeable difference is the emergence of a stellar cusp in a DM cusp.

The total mass of the nucleus remnants consisting of stars fallen into the centre can depend not only on DM density but also stellar distribution. If $R_{\rm h}$ is larger, stars have more extended distribution, therefore DF timescale becomes longer on average. Thus, a larger $R_{\rm h}$ leads a smaller fraction of stars to fall into the centre by DF. As a result, a less prominent stellar cusp would form in such an extended galaxy.

\section{$N$-body simulations}
\label{SimSec}
As I showed in Section \ref{analysis}, the analytic formulae are useful to estimate the magnitude of DF. The formulae, however, still ignore non-linear effects. For example, they assume DF as a corrective effect of two-body interactions and do not take into account orbital periodicity of particles or reaction of field particles. To address further the effect of DF using more realistic models, I perform $N$-body simulations in which the models are fully self-consistent, and DF drag force naturally arises as mutual interactions between particles.

\subsection{Settings}
The initial conditions of my $N$-body simulations are the same as the DM and stellar models (equation \ref{density} and \ref{star}) with the parameters used in Fig. \ref{OI_Hernq}: $r_{\rm s}=125~{\rm pc}$, $\sigma_{\rm h}=1.5~{\rm km~s^{-1}}$ (ignoring stellar potential) and $b=20~{\rm pc}$. The total stellar mass is set to $M_\star=2500~{\rm M_\odot}$, and a mass of a single stellar particle is $m_\star=0.5~{\rm M_\odot}$, i.e. the number of stellar particles is $N_\star=5000$. A stellar particle has a softening length of $\epsilon_\star=0.1~{\rm pc}$. Velocity distribution is given by Eddington's formula taking into account the total potential of the DM and the stars. Although the actual LOSV dispersion of stars inside $R_{\rm h}$ is slightly higher than $1.5~{\rm km~s^{-1}}$ because of self-gravity of the stars, the increase is only a few per cent. Although every single star is resolved with a point-mass particle, the interactions between stars in the simulations are still collisionless (see Appendix \ref{collision}).

DF can arise if $m_\star\gg m_{\rm DM}$, where $m_{\rm DM}$ is a mass of a DM particle in simulations. Since $m_\star=0.5~{\rm M_\odot}$ in my simulations, $m_{\rm DM}$ should be $\lsim0.05~{\rm M_\odot}$. Achieving such a high resolution requires approximately $2.6$ and $16.0\times10^7$ particles for the cuspy and the cored DM haloes. To lighten the heavy burden of the $N$-body computations, I employ an orbit-dependent refinement method for a multi-mass spherical model proposed by \citet{zms:08}. This method divides a DM halo into $i$ shells and the central sphere (the zeroth shell). Basically, each shell is resolved into DM particles with each mass resolution $m_{{\rm DM},i}$ and softening length $\epsilon_i$ (see Table \ref{multimassmodel}). After assigning a DM particle in the $i$-th shell its initial position and velocity and computing its pericentre distance in the fixed potential, if the pericentre intrudes into the inner $j$-th shell, the particle is split into $m_{{\rm DM},i}/m_{{\rm DM},j}$ particles with the mass $m_{{\rm DM},j}$ and the softening length $\epsilon_j$.\footnote{Therefore, the mass ratio $m_{{\rm DM},i}/m_{{\rm DM},j}$ has to be a natural number.} The split particles are distributed on random positions while keeping the initial radius of the parent particle, and directions of their tangential velocities are randomly reassigned while keeping the initial radial velocity and the kinematic energy of the parent particle. This refinement method can, by a substantial factor, reduce the computational run time by decreasing the number of DM particles in outer regions that are not important to this study, while preventing the outer particles with larger masses from entering the innermost region resolved with the smallest particle mass. After the refinement, $1.53$ and $6.88\times10^7$ particles are required to represent the cuspy and the cored DM haloes.
\begin{table}
  \caption{The multi-shell structures for the refinement method in my $N$-body simulations. From left to right, numbers of the shells, radial ranges of the shells, basic mass resolutions and softening lengths of the DM particles in the shells. By the orbit-based method for refinement, not all particles in the $i$-th shell have $m_{{\rm DM},i}$ although the central sphere ($i=0$) consists of the finest-resolution particles of $m_{{\rm DM},0}$.}
  \label{multimassmodel}
  $$ 
  \begin{tabular}{ccccc}
    \hline
     $i$-th shell & range & $m_{{\rm DM},i}~[{\rm M_\odot}]$ & $\epsilon_i~[{\rm pc}]$  & \\
    \hline
    0 & $r<r_{\rm s}$  & $0.05$ & $0.1$ & \\
    1 & $r_{\rm s}<r<2r_{\rm s}$  & $0.1$ & $0.14$ & \\
    2 & $2r_{\rm s}<r<3r_{\rm s}$  & $0.2$ & $0.2$ & \\
    3 & $3r_{\rm s}<r<4r_{\rm s}$  & $0.4$ &$0.28$& \\
    4 & $r>4r_{\rm s}$  & $0.8$ & $0.4$ & \\
    \hline
  \end{tabular}
  $$ 
\end{table}

I use a simulation code {\tt ASURA} \citep{sdk:08,sdk:09,sm:09,sm:10,sm:13},\footnote{{\tt ASURA} is an $N$-body/smoothed particle hydrodynamics (SPH) code although this study only uses the $N$-body part.} in which a symmetric form of a Plummer softening kernel \citep{sm:12}, a parallel tree method with an computational accelerator GRAPE \citep[GRAvity PipE,][]{m:04,tyn:13} and the second-order leap-frog integrator with individual timesteps are used. The number of stellar particles, $N_\star=5000$, in my simulations may be too small to obtain a statistically certain density profile. To reinforce this point, I perform ten runs with the same initial condition but different random-number seeds.

\subsection{Results}
\label{Results}
\subsubsection{Evolution of the stellar density profiles}
\label{Nbodystellar}
I obtain three surface density profiles observed from perpendicular angles for each of the ten runs. Then, I compute a stacking of the thirty profiles of stellar surface density at the same time $t$ for each case of the cuspy and the cored halo. The centre of the stellar distribution is defined to be the median position among all stellar particles in each snapshot.
\label{simulation}
\begin{figure}
  \includegraphics[width=\hsize]{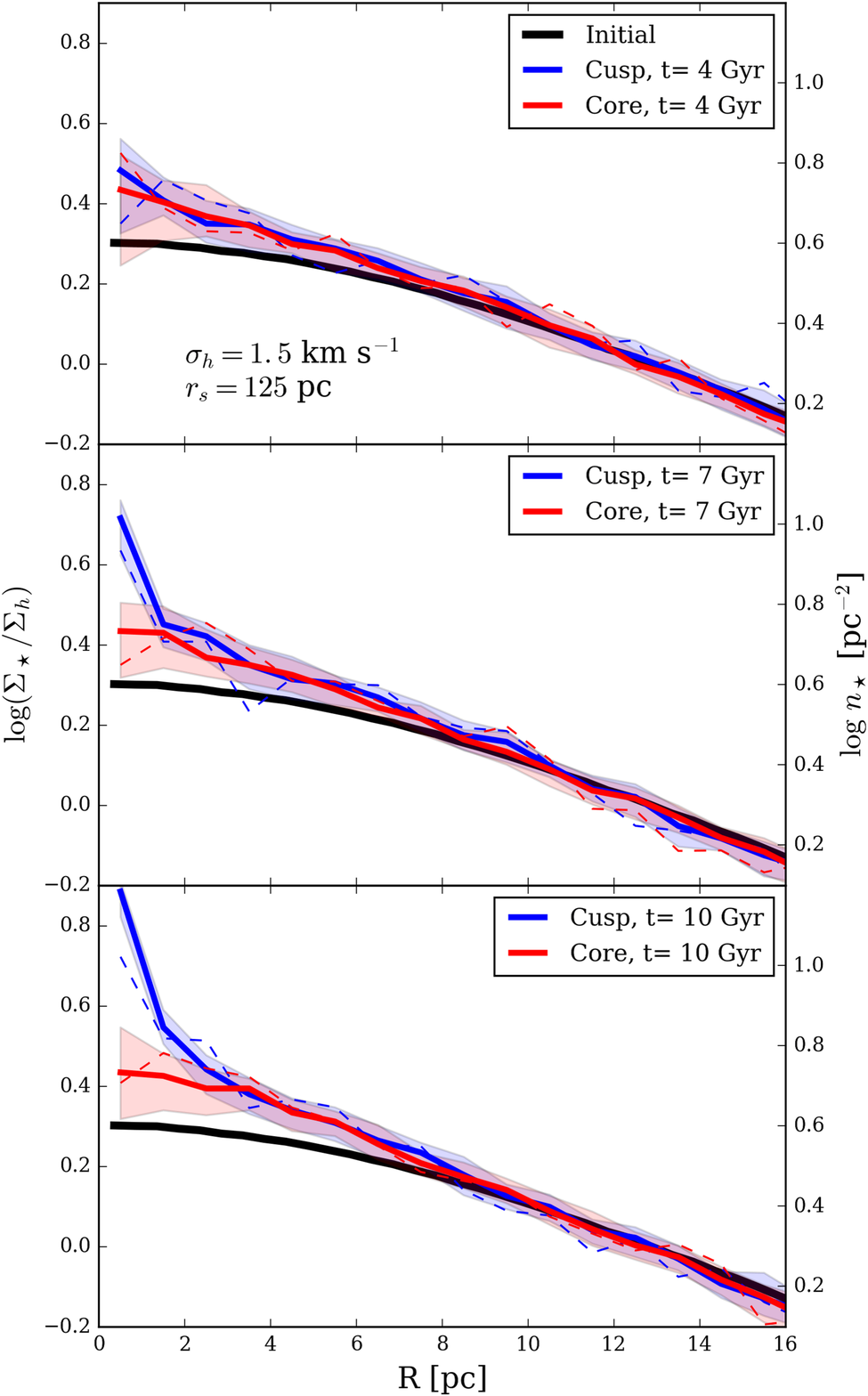}
  \caption{Time-evolution of stellar surface density profiles in the $N$-body simulations. The blue and red colours correspond to the runs of the cuspy and the cored DM halo models. The solid lines indicate the median values of the stackings of the thirty profiles (see the main text), and the shaded regions are the ranges of upper and lower $1\sigma$-deviations. The thin dashed lines is an example of a single profile chosen randomly in each time and DM model. The black solid lines are the initial state of the stellar density profiles. The left and right ordinate indicate surface densities normalised by $\Sigma_{\rm h}\equiv M_\star/(2\pi R_{\rm h}^2)=0.99~{\rm M_\odot~pc^{-2}}$ and number densities of stars, respectively.}
  \label{Nbody}
\end{figure}

Fig. \ref{Nbody} shows the stackings of stellar surface density profiles in the cuspy (blue) and the cored (red) DM haloes at $t=4$, $7$ and $10~{\rm Gyr}$. The shaded regions indicate the ranges of upper and lower $1\sigma$-deviations of the stackings. In the DM cusp, the stellar density profile clearly demonstrates the emergence of a stellar cusp after $t=7~{\rm Gyr}$; the density slope becomes remarkably steeper in $R\lsim2~{\rm pc}$ than that in $R\gsim2~{\rm pc}$. On the other hand, such a stellar cusp does not emerge in the cored DM halo although the outer density slope of the stars in $R\gsim2~{\rm pc}$ is similar to that in the cuspy DM halo, which is nearly exponential with radius. From this result, it can  be seen that a DM cusp can generate a stellar cusp by DF even if the stellar density profile is initially flat. In the $N$-body simulations, the stellar cusps have masses of $33.8^{+4.5}_{-5.3}~{\rm M_\odot}$ within $R<2~{\rm pc}$, which corresponds to $1.4$ per cent of the total stellar mass. Additionally, the difference between the two cases is significant in spite of the same $\sigma_{\rm h}$, which means that the two halo models are considered to be similar in observations. The emergence and the absence of stellar cusps in the cuspy and cored DM are approximately consistent with the results of my orbital integration models using the analytic DF formulae (Fig. \ref{OI_Hernq} and \ref{OI_others}).

\label{simulation}
\begin{figure}
  \includegraphics[width=\hsize]{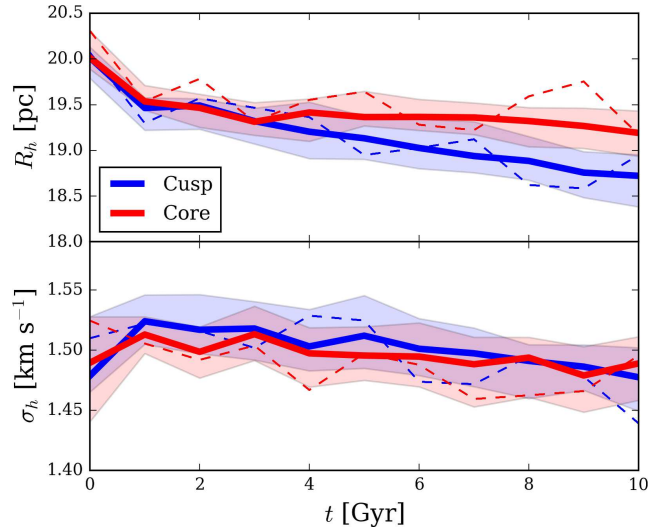}
  \caption{Time-evolution of $R_{\rm h}$ (top) and $\sigma_{\rm h}$ (bottom) in the $N$-body simulations. The solid lines indicate the median values of the stackings of the thirty stellar profiles, and the shaded regions are the ranges of $\pm1\sigma$-deviations. The thin dashed line is an example chosen randomly in each DM model.}
  \label{Rh_LOSVD}
\end{figure}
Fig. \ref{Rh_LOSVD} shows the evolution of $R_{\rm h}$ and $\sigma_{\rm h}$ during the $N$-body simulations. Stellar half-mass radii $R_{\rm h}$ decrease slightly; by $\simeq0.5$ and $1~{\rm pc}$ in the cored and cuspy DM. LOSV dispersions $\sigma_{\rm h}$ with in $R_{\rm h}$ are almost constant even after the emergence of the stellar cusps. This means that DF is not effective for most of stars around the half-mass radius although the central regions in $r\ll R_{\rm h}$ are significantly affected.

\subsubsection{Evolution of the DM density profiles}
As I showed above, DM exerts DF on stars and can cause a low-mass compact UFD to have a stellar cusp if it has a cuspy DM halo. On the other hand, the DM particles can be kinematically heated by the stars spiraling into the centre, as the reaction of DF. Previous studies have shown that a DM cusp can be disrupted or made shallower by objects spiraling into the centre \citep[e.g.][]{gmr:10,cdw:11,ac:17}. \citet{is:11} also demonstrated that a DM cusp disrupted by infalling objects can be revived if a central remnant of the infalling objects is sufficiently massive. Hence, it is also interesting to look into evolution of the dark matter density profiles in my $N$-body simulations, i.e. whether the cuspy halo is still cuspy or cored after the creation of the stellar cusp.
 
\begin{figure}
  \includegraphics[width=\hsize]{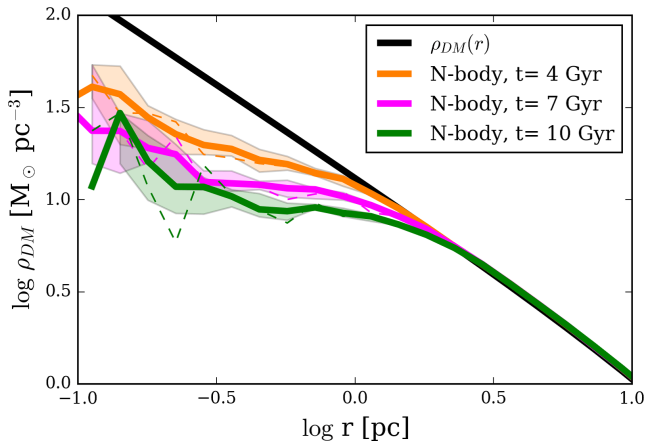}
  \caption{Profiles of spatial DM densities in the $N$-body simulations of the cuspy halo model at $t=4$, $7$ and $10~{\rm Gyr}$. The solid lines indicate the median values of the stacked profiles of DM, and the shaded regions cover $\pm1\sigma$-deviations among the ten runs. The thin dashed line is an example of a single run chosen randomly in each time. The black line delineates the analytic model of the DM distribution (equation \ref{density} with $\gamma=1$).}
  \label{DMcusp}
\end{figure}

Fig. \ref{DMcusp} shows DM density profiles in my $N$-body simulations of the cuspy halo model, in which I make a stacking of the ten runs. Here, the halo centre is defined to be a position of the particle that has the highest DM density in each snapshot. I use a method like SPH to compute the local DM densities for the centering; a cubic spline kernel is applied to 128 neighbouring DM particles. The Figure indicates that the DM cusp in the initial state is significantly weakened in $r\lsim1~{\rm pc}$ at $t=4~{\rm Gyr}$ (orange). Eventually, the initial DM cusp is turned into a core extending to $r\simeq2~{\rm pc}$ at $t=10~{\rm Gyr}$ (green). This result means that the central DM is kinematically heated by infalling stars, and the DM density is decreased in $r\lsim2~{\rm pc}$. The size of this region where DM is affected is consistent with the size of the stellar cusp in the $N$-body simulations (Fig. \ref{Nbody}). Since the softening length of DM particles is $\epsilon_0=0.1~{\rm pc}$, the peaks of DM densities at $r\simeq0.1~{\rm pc}$ may be transient fluctuation.

From the consistency of the sizes between the stellar cusps and the DM cores created, the size of a stellar cusp may be regulated by a DM density profile flattened by infalling stars. If this is the case, a larger number of stars falling into the centre can create a larger DM core and a broader stellar cusp. In the central region of a cuspy DM halo, the significance of DF basically depends on $\sigma_{\rm h}$\footnote{$F_{\rm DF}$ is almost independent from $r_{\rm s}$ in the \citeauthor{prg:16} formula for cuspy haloes (Fig. \ref{CF_comp}).}. In addition, the number of stars in the central region where stars can reach the centre by DF within $\sim10~{\rm Gyr}$ depends on the initial stellar distribution, i.e. $M_\star$ and $R_{\rm h}$.

\label{NbodyDM}

\section{Discussion and summary}
\label{Discussion}
\subsection{Summary and interpretation of the results}
\label{interp}
As I showed in Section \ref{analysis} and \ref{SimSec}, DF on stars against DM can largely alter the stellar density distribution if the galaxy has cuspy DM distribution and a compact stellar component having $R_{\rm h}\simeq20~{\rm pc}$ and $\sigma_{\rm h}\lsim3~{\rm km~s^{-1}}$. The most important result obtained from my $N$-body simulations is that the DF by the DM cusp can arouse emergence of a stellar cusp in the galactic centre $\lsim2~{\rm pc}$. On the other hand, if a DM halo has a core, DF is not efficient to generate such a stellar cusp, in spite of the same $\sigma_{\rm h}$, although stellar density can be affected and increase slightly in a wide range $\lsim10~{\rm pc}$.

The results mentioned above can be explained by the differences of density and velocity dispersion between a cuspy and a cored DM haloes. According to the analytic formulae, DF becomes stronger when a background density is higher and a velocity dispersion is lower. In a cuspy halo, DM density increases toward the centre, and velocity dispersion decreases (see Fig. \ref{velocities}), therefore DF becomes stronger towards the centre. In this case, orbital shrinkage by DF brings a star to an inner region where DF is even stronger: `DF shrivelling instability' \citep{h:16}. On the other hand, in a cored halo, density and velocity dispersion are nearly constant in the central region, therefore DF drag force is approximately independent of radius; it actually decreases gently towards the centre (Fig. \ref{CF_comp}). This means that a cored halo is relatively stable against the DF shrivelling of stars.

Interpretation of the above results should be considered carefully. It should be noted that presence of a stellar cusp in a low-mass compact UFD is not necessarily evidence to prove a DM cusp. It is because we do not know the initial condition of the stellar density; a galaxy can create a stellar cusp at its birth even if its DM halo has a core. It could be said, however, that it is inevitable to have a stellar cusp if a low-mass compact UFD has a DM cusp. In other words, if a low-mass compact UFD is observed to have no stellar cusp and be old enough, it suggests that its DM halo would have a large core with a size of $r_{\rm s}\gsim100~{\rm pc}$. I discuss UFDs in current observations on this point in Section \ref{obs}.

\subsection{The analytic formulae vs $N$-body simulations}
It is interesting to compare the results of the analytic formulae with those of $N$-body simulations although it is not the main purpose of this study. Early studies using numerical simulations have discussed that the Chandrasekhar formula assuming Maxwellian velocity distribution and a invariant $\Lambda$ can give a quite accurate estimate of DF force in various cases \citep[e.g.][]{lh:83,bva:87}. It was also reported, however, that the formula can be inaccurate in some specific cases; analyses and $N$-body simulations have shown that DF can be enhanced around a constant-density core, and then suppressed in the core \citep[e.g.][]{gmr:06,i:09,gmr:10,ac:14,pgr:15,prg:16}. These phenomena are inconsistent with predictions by the simplified Chandrasekhar formula. Various physical mechanisms of the deviation from the analytic formula have been proposed: orbital resonance between a massive and field particles \citep{i:11,zk:16}, coherent velocity field among particles \citep{rgm:06}, a non-Maxwellian velocity distribution \citep{sls:16,pgr:15,jkb:11}, decrease of low-velocity particles \citep{am:12,prg:16,da:16} and inhomogeneity of background density and a variable $\Lambda$ \citep{jp:05}. 

In the case of a DM cusp, the two analytic formulae predict similar DF force in Fig. \ref{CF_comp}, and the results of my orbital integration models are qualitatively consistent with the $N$-body simulations. However, the stellar cusp in my $N$-body simulations have the size of $R\simeq2~{\rm pc}$, and it could be attributed to the weakened DM cusp shown in Fig. \ref{DMcusp}, where DF is weakened. In addition, the DM density centre is not necessarily be fixed onto the stellar centre in the simulations, and the slippage of the centres can broaden the stellar cusp. Therefore, the broadness of the stellar cusp in the simulations does not necessarily mean inaccuracy of the DF modellings. However, the stellar density slopes outside the stellar cusp is steeper in the orbital integration models. Moreover,  in my $N$-body simulations, the total mass of the stellar cusps within $R=2~{\rm pc}$ are about four times smaller than those predicted by the analytic DF modellings. On these points, both analytic formulae would be overestimating DF in the cuspy haloes.

In the case of a DM core, on the other hand, DF cannot generate a stellar cusp in either the orbital integration or the $N$-body models. However, there is a difference worthy of special mention: the simplified Chandrasekhar formula hardly changes the stellar density slopes, whereas the $N$-body simulations show significant increase of the stellar densities in a wide range of $R\lsim10~{\rm pc}$ (Fig. \ref{Nbody}). This result shows that the Chandrasekhar formula assuming Maxwellian distribution and a invariable $\Lambda$ underestimates DF in the cored halo despite that the classical formula overestimates in the cuspy halo. It is noteworthy that using the Petts et al. formula with their tidal-stalling model can dramatically improve the reproducibility of DF effect in the cored haloes, and the result of the stellar density profile is almost consistent with the $N$-body simulations (see the bottom panels of  Fig. \ref{OI_Hernq} and \ref{Nbody}). Thus, the DF modelling proposed by \citet{prg:16} seems to be more accurate than the simplified Chandrasekhar formula although it may not be perfect yet in a DM cusp. Although it is beyond the scope of this study to investigate the physical reasons of the differences between the analytic formulae and my $N$-body simulations, I consider that the $N$-body simulations would be physically more credible than the analytic models.

\subsection{Comparison with observations}
\label{obs}
Here, I discuss the validity of my models of low-mass compact UFDs and the results in comparison with current observations. First, it is still very difficult or impossible to determine masses and sizes of DM haloes of UFDs with accuracy in current observations. I have to note, therefore, that the parameters in my DM halo models might be arbitrary. Recent observational studies have argued that galaxies have the universal DM surface density, $\mu_{\rm DM}\equiv\rho_{\rm DM,0}r_{\rm s}$, over quite a wide range of luminosity when they are assumed to have cored DM haloes \citep{sma:08,hc:15,kf:16}.\footnote{This universality can be explained by assuming the Faber-Jackson law for DM haloes \citep{kf:16}.} \citet{d:09} and \citet{kf:16} derived $\mu_{\rm DM}=140^{+80}_{-30}$ and $70\pm4~{\rm M_\odot~pc^{-2}}$ from their galaxy samples including some satellite galaxies of the Milky Way. Although it has to be noted that they assumed different models for their cored haloes, my cored DM models of equation (\ref{density}) have $\mu_{\rm DM}=122$ and $176~{\rm M_\odot~pc^{-2}}$ for $r_{\rm s}=125$ and $250~{\rm pc}$, respectively, when $\sigma_{\rm h}=1.5~{\rm km~s^{-1}}$. Accordingly, my cored halo model with $r_{\rm s}=125$--$250~{\rm pc}$ and $\sigma_{\rm h}=1.5~{\rm km~s^{-1}}$ would be consistent with the observed universality of $\mu_{\rm DM}$ if it is extrapolated to extremely low-mass galaxies. 

For my stellar model, I assume the uniform stellar mass of $m_\star=0.5~{\rm M_\odot}$. However, of course, stars generally have different masses according to their initial mass function and stellar evolution. Although stellar scattering by massive stars would not be efficient since encounters between stars are expected to be rare in a low-mass compact UFD (see Appendix \ref{collision}), mass segregation does occur on the same timescale as DF because of mass-dependence of DF. Because more massive stars sink faster into the centre of a DM cusp, a stellar cusp would mainly consist of massive stars.\footnote{The massive objects are
stellar remnants. Even the typical mass of White Dwarfs is larger than $0.5~{\rm M_\odot}$.} The massive objects in the stellar cusp could be a heating source of less massive stars around it and prevent the less massive stars from falling into the centre. Thus, I have to note that my $N$-body simulations lack this effect.

The best UFD that is the most similar to my $N$-body model ($R_{\rm h}\simeq20~{\rm pc}$, $\sigma_{\rm h}=1.5~{\rm km~s^{-1}}$ and $\log(M_\star/M_\odot)=3.4$) would be Draco II, which has $R_{\rm h}\sim19^{+8}_{-6}~{\rm pc}$, $\sigma_{\rm h}=2.9\pm2.1~{\rm km~s^{-1}}$, the total  luminosity $\log(L_{\rm V}/L_\odot)=3.1\pm0.3$ and an age  $\sim12~{\rm Gyr}$ \citep{l:15,m:16}. Although the most probable value of $\sigma_{\rm h}$ observed is nearly twice higher than that in my model, it is within the error range. If I adopt a stellar mass-to-luminosity ratio $M/L_{\rm V}=2$--$3~{M_\odot/L_\odot}$ for a metal-poor system of $12~{\rm Gyr}$ from a simple stellar population model of \citet{m:05}, the stellar mass of Draco II is approximately $\log(M_\star/M_\odot)=3.1$--$3.9$. At the Heliocentric distance of Draco II, $20\pm3~{\rm kpc}$ \citep{l:15}, the size of a stellar cusp expected from my $N$-body simulations, $\simeq2~{\rm pc}$, corresponds to $\simeq0.3~{\rm arcmin}$. Unfortunately, the cusp region is smaller than the size of the innermost bin of a stellar surface density profile shown in fig. 3 of \citet{l:15}, therefore it might be still challenging for the current observations to detect a stellar cusp in Draco II even if it is present. In addition, the observed number of stars belonging to Draco II may be too small to excavate a stellar cusp in the current observations. Although DM density profiles of UFDs may differ from one another even if they have the same sizes and LOSV dispersions, it could improve statistics for proving absence of stellar cusps to make a stacking like Fig. \ref{Nbody} among UFDs having similar and sufficiently small $R_{\rm h}$ and $\sigma_{\rm h}$. Because of the faintness of UFDs as small as Draco II, observations are limited to the close distance from the solar system: $d\lsim30~{\rm kpc}$. It could be expected, however, that future observations will explore vaster regions to discover such faint UFDs. 

\section*{Acknowledgments}
The author thanks the referee for his/her useful comments that helped improve
the article greatly, and Takayuki R. Saitoh for kindly providing the simulation code {\tt ASURA}. This study was supported by World Premier International Research Center Initiative (WPI), MEXT, Japan and CREST, JST. The numerical computations presented in this paper were carried out on Cray XC30 at Center for Computational Astrophysics, National Astronomical Observatory of Japan.


\appendix
\section{Analytic solution of interaction intensity}
\label{anaJ}
Equation (\ref{Petts_J}) integrates intensity of interactions with DM particles over possible relative velocities and impact parameters. \citet{prg:16} have mentioned that their formula (equation \ref{Petts_formula}) containing $J(v_{\rm DM})$ requires a double integral which is quite expensive in numerical computations. For the sake of practical use of their formula, here I find the analytical solution of equation (\ref{Petts_J}).

By letting $A\equiv v_\star^2-v_{\rm DM}^2$ and $B=b_{\rm max}/G^2m_\star^2$, the indefinite integral of equation (\ref{Petts_J}) can be obtained as,

\begin{equation}
  j(V) = \int\left(1+\frac{A}{V^2}\right)\log\left(1+BV^4\right)~\textrm{d}V
\label{indefJ}
\end{equation}
\begin{equation}
\;\;\;\;\;\;\;\:= \left(V-\frac{A}{V}\right)\log\left(BV^4+1\right) -4V + \frac{I_{\rm 1} + {\rm Im}(I_{\rm 2})}{\sqrt{2}B^{1/4}} +C,
\end{equation}
where $C$ is an integration constant, and 
\begin{equation}
I_{\rm 1}(V) = (A\sqrt{B}-1)\log\left(\frac{\sqrt{B}V^2+1-\sqrt{2}B^{1/4}V}{\sqrt{B}V^2+1+\sqrt{2}B^{1/4}V}\right)
\end{equation}
\begin{equation}
I_{\rm 2}(V) = (A\sqrt{B}+1)\log\left(\frac{\sqrt{B}V^2-1-\sqrt{2}iB^{1/4}V}{\sqrt{B}V^2-1+\sqrt{2}iB^{1/4}V}\right).
\end{equation}
Furthermore, ${\rm Im}(I_{\rm 2})$ is transformed as follows,\footnote{${\rm Im}[\log(x+iy)]=\arctan(y/x)$ \citep{grj:07}.}
\begin{equation}
{\rm Im}(I_{\rm 2}) = -2(A\sqrt{B}+1)\arctan\left(\frac{\sqrt{2}B^{1/4}V}{\sqrt{B}V^2-1}\right).
\end{equation}
Eventually, the definite integral, equation (\ref{Petts_J}), is
\begin{equation}
J(v_{\rm DM}) = j(v_\star+v_{\rm DM}) - j(|v_\star-v_{\rm DM}|)
\end{equation}
Thus, since equation (\ref{Petts_J}) can be solved analytically, the equation (\ref{Petts_formula}) actually does not require a double integration but a single integration in the numerical computations.

\section{Collisionlessness of the simulations}
\label{collision}
In my $N$-body simulations, every single star is resolved with a point-mass particle, and I should ascertain whether the gravitational interactions between the stellar particles are collisional or collisionless. Perturbations on a star by the others can be approximated as,
\begin{equation}
  \Delta v_\perp^2 \simeq 8N_\star\left(\frac{Gm_\star}{r_\star v_\star}\right)^2\ln\Lambda_\star,
\end{equation}
where 
\begin{equation}
 \Lambda_\star\equiv\frac{b_{\star,{\rm max}}}{b_{\star,{\rm min}}}\sim\frac{v_\star^2r_\star}{Gm_\star},
\end{equation}
where $b_{\star,{\rm min}}\sim Gm_\star/v_\star^2$,  $b_{\star,{\rm max}}\sim r_\star$. Here, using a stellar mass fraction $f_\star$, $v_\star^2\sim GN_\star m_\star/(r_\star f_\star)$, and the number of crossings that are required for dynamical relaxation is  
\begin{equation}
 n_{\rm relax}\equiv\frac{v_\star^2}{\Delta v_\perp^2}=\frac{N_\star/f_\star^2}{8\ln\left(N_\star/f_\star\right)}.
\end{equation}
$f_\star\simeq0.03$ within $r_\star$ in both of the cuspy and the cored haloes. In the initial settings of my simulations, $n_{\rm relax}=5.8\times10^5$. The crossing time is $t_{\rm cross}\sim r_\star/v_\star\simeq r_\star/\sigma_{\rm h}=13.0~{\rm Myr}$. Eventually, I estimate the relaxation timescales to be approximately $t_{\rm relax}\equiv n_{\rm relax}\times t_{\rm cross}\sim10^3~{\rm Gyr}$ which are significantly longer then the age of the Universe: $\sim10~{\rm Gyr}$. Therefore, the stellar interactions in my $N$-body simulations are regarded to be collisionless.

\end{document}